\documentclass[a4paper,12pt]{article}

\usepackage{graphicx}
\usepackage{epsfig}
\usepackage{amsfonts,amsmath}
\usepackage{amssymb}
\usepackage{epsf}
\usepackage[hidelinks]{hyperref}
\usepackage{authblk}

\usepackage{comment}

\usepackage{color}
\usepackage[normalem]{ulem}
\usepackage[dvipsnames]{xcolor}
\usepackage{cite}

\usepackage[utf8]{inputenc}
\usepackage[english]{babel}

\usepackage{graphicx}
\usepackage{subcaption}
\usepackage{parskip}

\usepackage{float}
\usepackage{booktabs}
\usepackage{makecell}
\usepackage{multirow}
\usepackage{soul}

\usepackage[figurename=Fig.]{caption}
\usepackage[tablename=Tab.]{caption}
\usepackage[font=small]{caption}

\usepackage{amsmath}
\usepackage{amssymb}
\usepackage{amsfonts}
\usepackage{mathtools}

\usepackage{upgreek}
\usepackage{xcolor}
\usepackage[normalem]{ulem}

\title{Boson Stars surrounded by Polish Doughnuts in Scalar-Tensor Theory}

\author[1,2]{Kristian Gjorgjieski
\thanks{\href{mailto:kristian.gjorgjieski@uol.de}{kristian.gjorgjieski@uol.de}}}
\author[1]{Maxim Rose
\thanks{\href{mailto:maxim.rose@uni-oldenburg.de}{maxim.rose@uni-oldenburg.de}}}
\author[1]{Burkhard Kleihaus 
\thanks{\href{mailto:b.kleihaus@uni-oldenburg.de}
{b.kleihaus@uni-oldenburg.de}}}
\author[1]{\\ Jutta Kunz
\thanks{\href{mailto:jutta.kunz@uni-oldenburg.de}
{jutta.kunz@uni-oldenburg.de}}} 
\author[3]{Petya Nedkova
\thanks{\href{mailto:pnedkova@phys.uni-sofia.bg}
{pnedkova@phys.uni-sofia.bg}}}

\affil[1]{Institute of Physics, University of Oldenburg, Mailbox 2503,
D-26111 Oldenburg, Germany}
\affil[2]{Campus Institute Data Science, University of Goettingen, 37077 Goettingen}
\affil[3]{Department of Theoretical Physics, Sofia University, Sofia 1164, Bulgaria}
\date{\today}

\begin{document}

\maketitle

\begin{abstract}
We investigate thick accretion disks (Polish Doughnuts) around rotating self-inter\-acting boson stars in general relativity and scalar–tensor theories, focusing on spontaneously scalarized solutions and their general relativistic counterparts. Using equilibrium models with constant specific angular momentum, we analyze disk structures across the parameter space, with emphasis on the phase transition between GR and scalarized configurations. We find that scalarization induces qualitative changes in the spacetime that significantly affect disk morphology. In particular, scalarized boson stars can lack innermost circular orbits, allowing stable motion down to the center and enabling highly compact, quasi-spherical disks. For the most massive scalarized solutions, a non-monotonic angular momentum profile further permits two-centered disk configurations connected by a cusp.
Overall, disks around scalarized boson stars are more compact and more strongly bound than those in general relativity, highlighting distinctive features that may serve as observational signatures of alternative gravity theories in the strong-field regime.
\end{abstract}

\section{Introduction}

In the strong gravity regime, General Relativity (GR) has so far passed all tests within observational accuracy \cite{Will:2005va,Will:2018bme}, ranging from measuring the orbits of stars around supermassive black holes \cite{Genzel:2024vou,GRAVITY:2023avo,GRAVITY:2024tth} 
and imaging their shadows \cite{EventHorizonTelescope:2019dse,EventHorizonTelescope:2022wkp}
to the detection of gravitational waves from the merger of black holes and neutron stars \cite{LIGOScientific:2016aoc,LIGOScientific:2017vwq,Cahillane:2022pqm},
and the analyses of the highly accurate pulses from binary pulsars \cite{Freire:2024adf}.
Still, there are compelling reasons to venture beyond GR and study the distinctive observational signals in the strong gravity sector arising in alternative theories of gravity \cite{Faraoni:2010pgm,Berti:2015itd,CANTATA:2021ktz}.

Accretion disks are accumulations of matter that form around compact objects like black holes and neutron stars. They can be viewed as fundamental natural laboratories for high-energy astrophysics, which encode information about the central object around which they form \cite{Abramowicz:1994ih}. Their spectra, variability, and relativistic line profiles allow us to probe strong-gravity effects such as frame dragging, orbital dynamics, and potential deviations from general relativity. Besides that, they also constrain the physics of magnetic turbulence, angular momentum transport, and power some of the universe’s most energetic phenomena like astrophysical jets, quasars, and X-ray binaries. Their study is essential for understanding galaxy evolution and feedback processes \cite{Frank:2002,Kato:2008}. Moreover, also on small scales recent multi-wavelength and gravitational wave observations have demonstrated that detailed disk modeling is crucial for interpreting signals from tidal disruption events, black hole mergers and rapidly accreting neutron stars.

Here we consider thick accretion disks, that are also often referred to as Polish Doughnuts \cite{Abramowicz:1994ih}. In particular, their study is valuable, since they represent a regime of extreme astrophysical conditions that thin disk models cannot capture. These geometrically thick, radiation supported structures naturally arise in environments with very high accretion rates, such as those surrounding rapidly growing black holes or compact object mergers. The thick disk model provides closed form solutions, given the spacetime metric, which capture the main properties of possible accretion disk solutions, and enable thus a qualitative study of potentially observational signatures.

Especially when it comes to alternative theories of gravity, the study of accretion disks can be informative, as most of the new physics effectively only appears in the strong-field regime, which can not be tested in the laboratory or our solar system. The structure and stability of thick disks are highly sensitive to the underlying spacetime geometry, since predictions about orbital frequencies, innermost stable orbits, and energy extraction mechanisms can differ across gravitational theories. As such they provide an interesting testing ground in the strong-field regime to identify possible deviations from general relativity.

Here we consider scalar-tensor theories (STTs). STTs are a well-established extension of general relativity and have a long history \cite{Jordan:1949zz,Fierz:1956zz,Jordan:1959eg,Brans:1961sx,Dicke:1961gz}. Besides the metric tensor, STTs feature an additional gravitational scalar field. The presence of this scalar field has led to severe constraints for some STTs (see e.g.~\cite{Will:2018bme,Freire:2024adf}).
In particular, STTs giving rise to the interesting phenomenon of matter induced spontaneous scalarization of neutron stars \cite{Damour:1993hw,Doneva:2013qva} are basically ruled out for a massless gravitational scalar field. When allowing for a small scalar mass, however, these constraints can be circumvented \cite{Ramazanoglu:2016kul,Yazadjiev:2016pcb}.

Matter induced spontaneous scalarization occurs also for boson stars \cite{Torres:1997np,Whinnett:1999sc,Alcubierre:2010ea,Ruiz:2012jt,Kleihaus:2015iea,Evstafyeva:2023kfg,Huang:2025dgc}. They arise naturally by assuming dark matter in the form of scalar or vector bosons, of which they are composed \cite{Jetzer:1991jr,Lee:1991ax,Schunck:2003kk,Liebling:2012fv}). Numerous features of boson stars have been addressed with respect to theoretical and astrophysical questions. Here, our focus will be on the study of thick accretion disks of spontaneously scalarized boson stars, employing a repulsive 4th-order potential for the self-interaction of the complex scalar field \cite{Colpi:1986ye}. Since these STTs retain the unscalarized GR solutions, we will also present a comparison between disks surrounding scalarized and GR boson stars.

The paper is organized as follows: In Section 2 we establish the theoretical framework by discussing STTs, (un)scalarized boson stars, and thick accretion disks. Afterwards, we present our results in Section 3 for thick tori around scalarized boson star and GR boson stars. In section 4 we compare their properties and conclude.

\section{Theoretical Framework}

\subsection{Scalar-tensor theories}

We start by briefly recalling STTs.
In the physical Jordan frame their action is given by
\begin{eqnarray} \label{JFA}
S &=& \frac{1}{16\pi G_{*}} \int d^4x \sqrt{-{\tilde
g}}\left({F(\Phi)\tilde R} - Z(\Phi){\tilde
g}^{\mu\nu}\partial_{\mu}\Phi
\partial_{\nu}\Phi   -2 W(\Phi) \right) \nonumber \\[2ex]
&+&
S_{m}\left[\Psi_{m};{\tilde g}_{\mu\nu}\right] .
\end{eqnarray}
It consists of the gravitational action with metric ${\tilde g}_{\mu\nu}$, Ricci scalar curvature ${\tilde R}$, gravitational scalar field $\Phi$, and the matter action $S_{m}\left[\Psi_{m};{\tilde g}_{\mu\nu}\right]$ with the complex scalar field $\Psi_m$. $G_{*}$ denotes the gravitational constant, the functions $F(\Phi)$ and $Z(\Phi)$
need to satisfy physical restrictions, namely $F(\Phi)>0$, (so that gravitons have positive energy) and
\begin{equation}
2F(\Phi)Z(\Phi) + 3[dF(\Phi)/d\Phi]^2 \ge 0,
\end{equation}
so that the kinetic energy of the scalar field is non-negative. In order not to violate the weak equivalence principle, the matter action depends only on the metric ${\tilde g}_{\mu\nu}$ and the boson field $\Psi_{m}$ is minimally coupled. The thick tori will be studied directly in the Jordan frame, but the construction of the boson stars is performed in the Einstein frame. Both frames are conformally related by
\begin{equation}\label {CONF1}
g_{\mu\nu} = F(\Phi){\tilde g}_{\mu\nu},
\end{equation}
where $g_{\mu\nu}$ denotes the metric in the Einstein frame.
Introducing the scalar field $\phi$ via
\begin{equation}\label {CONF2}
\left(\frac{d\phi}{d\Phi} \right)^2 = \frac{3}{4}\left(\frac{d\ln(F(\Phi))}{d\Phi } \right)^2 + \frac{Z(\Phi)}{2 F(\Phi)} ,
\end{equation}
and the functions
\begin{equation}\label{CONF3}
{\cal A}(\phi) = F^{-1/2}(\Phi) \, , \,\, 
2V(\phi) = W(\Phi)F^{-2}(\Phi) ,
\end{equation}
the action becomes in the Einstein frame
\begin{eqnarray}
S= \frac{1}{16\pi G_{*}}\int d^4x \sqrt{-g} \left(R -
2g^{\mu\nu}\partial_{\mu}\phi \partial_{\nu}\phi -
4V(\phi)\right)+ S_{m}[\Psi_{m}; {\cal A}^{2}(\phi)g_{\mu\nu}] ,
\end{eqnarray}
where the Ricci scalar $R$ is defined in terms of the metric $g_{\mu\nu}$.
Variation of this action leads to a coupled set of field equations in the Einstein frame. We employ here a coupling function ${\cal A}(\phi)$ that can give rise to spontaneous scalarization \cite{Damour:1993hw,Doneva:2013qva,Kleihaus:2015iea}
\begin{equation}
\ln {\cal A}(\phi) = \frac{1}{2} \beta \phi^2  ,
\label{funA}
\end{equation}
when the parameter $\beta$ is suitably chosen.
In order to evade observational constraints the gravitational scalar field should possess a small mass $m_\phi$\cite{Ramazanoglu:2016kul,Yazadjiev:2016pcb}, leading to the potential 
\begin{equation}
    V(\phi)= \frac{1}{2} m_\phi^2 \phi^2 .
\end{equation}

In the Einstein frame the matter action $S_{m}[\Psi_{m}; {\cal A}^{2}(\phi)g_{\mu\nu}]$ reads
\begin{equation}\label{Smat}
S_{m}[\Psi_{m}; {\cal A}^{2}(\phi)g_{\mu\nu}]
= - \int d^4x \sqrt{-g} 
\left[ \frac{1}{2} {\cal A}^2(\phi)  g^{\mu\nu}
\left( \Psi_{, \, \mu}^* \Psi_{, \, \nu} 
+ \Psi _ {, \, \nu}^* \Psi _{, \, \mu} \right) 
+ {\cal A}^4(\phi)  U( \left| \Psi \right|) \right].
\end{equation}
For the self-interaction potential $U( \left| \Psi \right|)$ of the complex boson field $\Psi$ we choose 
\begin{equation}\label{SmatU}
U( \left| \Psi \right|) =
m_b^2 \left| \Psi \right|^2 + \Lambda \left| \Psi \right|^4,
\end{equation}
where $m_b$ is the boson mass and $\Lambda$ is the coupling constant.

\subsection{Boson stars}
Rotating boson stars can be constructed by solving for the stationary axially symmetric line element 
\cite{Kleihaus:2000kg,Kleihaus:2004gm,Kleihaus:2015iea}
\begin{equation}
ds^2 = - f_0 dt^2 + \frac{f_1}{f_0} \left( f_2 \left[
dr^2 + r^2 d \theta^2 \right]
+ r^2 \sin^2 \theta \left[ d\varphi - f_3 dt \right]^2 \right) ,
\label{metric}
\end{equation}
with the metric functions $f_i(r,\theta)$, $i=0,...,3$.
The gravitational scalar field $\phi$ is parametrized as $\phi(r,\theta)$, and the complex boson field $\Psi$ is expressed as
\begin{equation}
\Psi = \psi(r,\theta) \, e^{i (\omega t +  n \varphi)} ,
\end{equation}
where $\psi$ is a real function, $\omega$ is the boson frequency, and $n$ is an integer ($n\ne 0$).

Requiring regularity at the origin and on the symmetry axis together with asymptotic flatness of the boson stars determines the boundary conditions
\begin{eqnarray}
& & \partial_r f_i(0,\theta) = 0 , \ \ \ i=0,1,2,3  \ ,  \ \ \
 \psi(0,\theta) = 0 \ , \ \ \ \partial_r \phi(0,\theta) = 0 , \\
& &  f_i(\infty,\theta) = 1, \ \ \ i=0,1,2 \ , \ \ \ f_3(\infty,\theta) =0 \ ,  \ \ \
 \psi(\infty,\theta) =\phi(\infty,\theta) = 0 , \\
& &  \partial_\theta f_i(r,0) = 0 , \ \ \ i=0,1,3 \ , \ \ \ f_2(r,0) =1 \ ,  \ \ \
 \psi(r,0) = 0 \ , \ \ \partial_\theta \phi(r,0) = 0 , \\
& &  \partial_\theta f_i(r,\pi/2) = 0 , \ \ \ i=0,1,2,3 \ ,  \ \ \
 \partial_\theta \psi(r,\pi/2) =  \partial_\theta \phi(r,\pi/2) = 0 \ .
 \end{eqnarray}
With the last set of boundary conditions we require also reflection symmetry with respect to the equatorial plane. The mass $M$ and the angular momentum $J$ of the boson stars are given by the asymptotic expansion of the metric functions $f_0$ and $f_3$,
\begin{eqnarray} 
f_0 &=& 1 -2 M/r + \mathcal{O}(r^{-2}), \\
f_3 &=&  2 J/r^3 + \mathcal{O}(r^{-4}). 
\end{eqnarray} 
These charges hold also in the Jordan frame, since the spacetime is asymptotically flat.

As detailed in \cite{Kleihaus:2015iea,Huang:2025dgc}, the coupled equations are solved subject to the above set of boundary conditions, making use of the professional package FIDISOL/CADSOL \cite{FIDISOL}.
The self-interaction coupling constant has the value $\Lambda=300$, the STT parameter is chosen as $\beta=-10$.
We employ the mass of the gravitational scalar $m_\varphi=0$, since a small but finite mass in the observationally allowed region (e.g.~$m_\varphi/m_b=10^{-3}$) has almost negligible effect on the spacetime properties \cite{Huang:2025dgc}.
All the rotating boson stars considered here possess the rotational number, $n=1$.
These rotating boson stars are axially symmetric and possess a torus-like energy density.
Thus their boson field assumes its maximum value on a ring in the equatorial plane \cite{Schunck:1996,Schunck:1996he,Ryan:1996nk,Yoshida:1997qf,Schunck:1999pm,Kleihaus:2005me,Kleihaus:2007vk}. Using the Planck mass $m_{Pl}$ we can express the mass $M$ and angular momentum $J$ of the boson stars in dimensionless units,
\begin{equation}
M_0 = m_{Pl}^2/m_b, ~ J_0= m_{Pl}^2/m_b^2.
\end{equation}

\begin{figure}[H]
\begin{center}
\vspace{-0.5cm}
\mbox{
\hspace*{-0.3cm}\includegraphics[height=.27\textheight, angle =0]{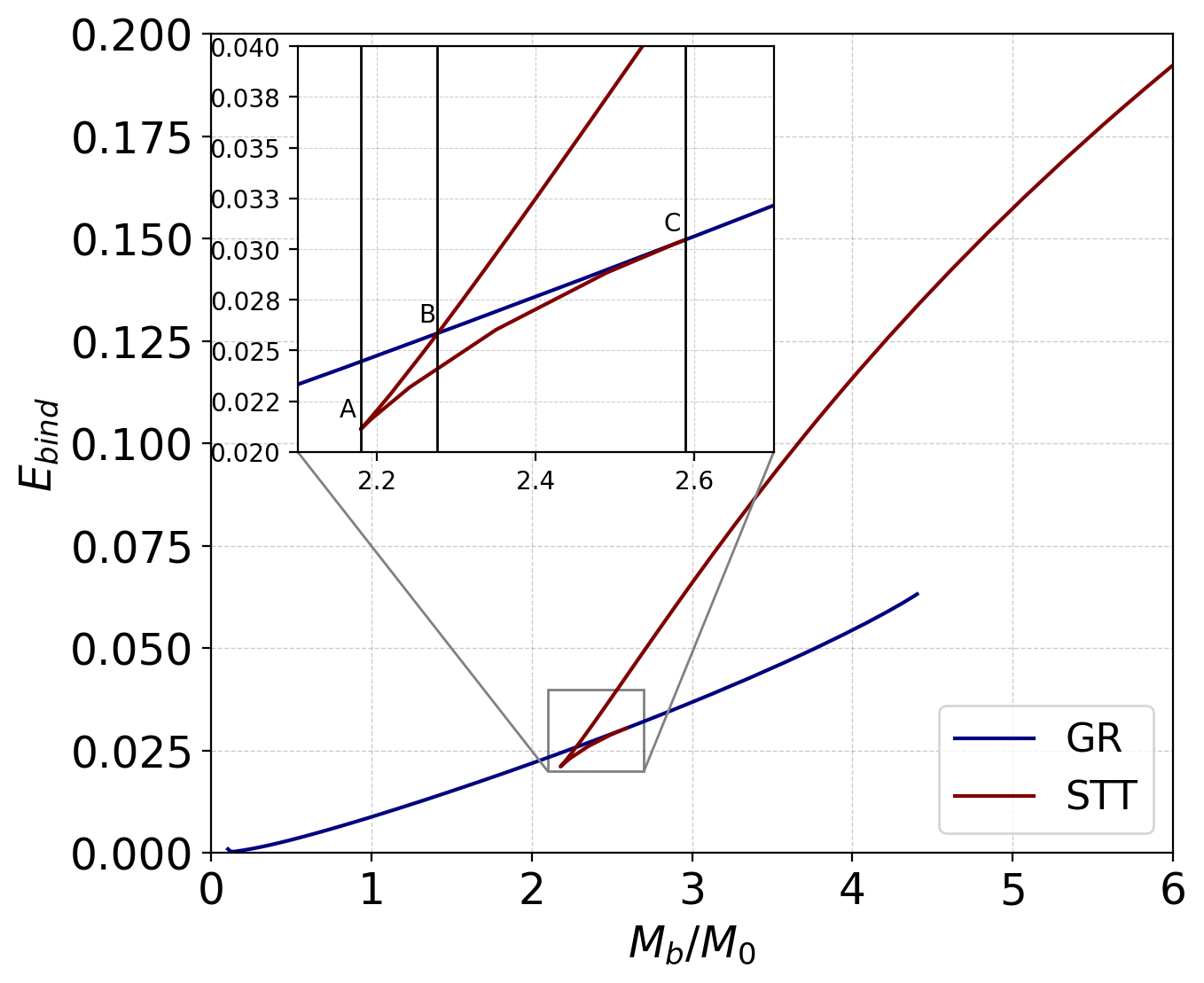}
\hspace*{-0.1cm}\includegraphics[height=.27\textheight, angle =0]{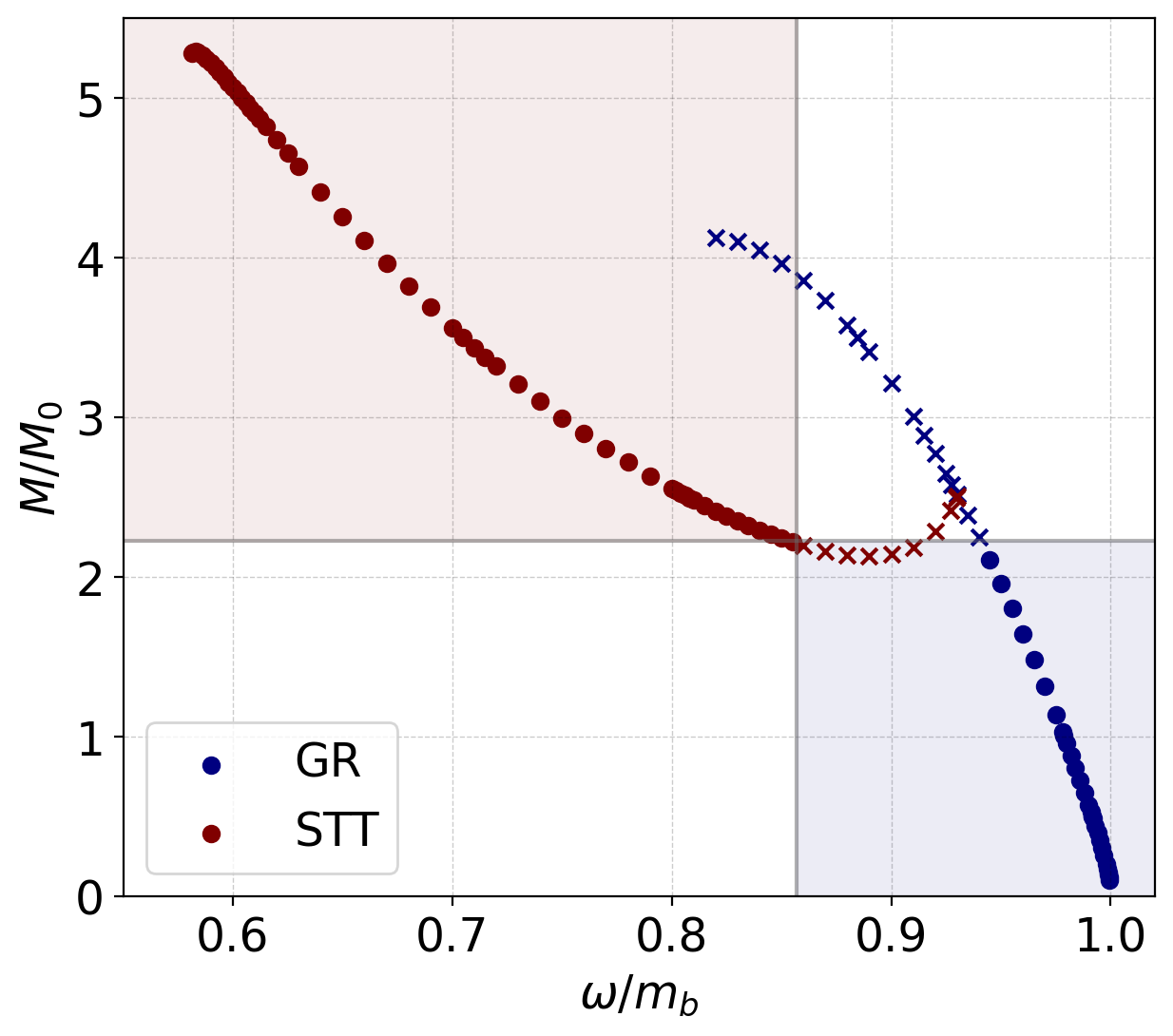} 
}
\end{center}
\vspace{-0.5cm}
\caption{Rotating boson stars: (a) shows the binding energy versus the scaled boson mass of the GR and STT solutions, with the inset illustrating an enlargement of the region where a phase transition occurs between GR and STT boson stars (point B). Points A and C mark the emergence of STT solutions. (b) showcases the dimensionless mass versus the boson frequency  of the solutions. The red and blue background color mark the parameter regions in which the STT/GR boson stars have a higher binding energy than their GR/STT counterparts, respectively. The intersection of both straight lines marks point B from figure (a), where the phase transition is occurring.}
\label{fig1}
\end{figure}

We exhibit the sets of unscalarized (GR) and scalarized (STT, $\beta=-10$) boson stars in Fig. \ref{fig1}. The subfigure (a) shows hereby the dimensionless binding energy $E_\text{bind}=(M_b-M)/M_b$ (where $M_b$ is the total boson mass) versus the dimensionless boson mass.
The STT boson stars bifurcate at the critical point $C$ from the GR ones. Beyond the bifurcation, the mass of the scalarized boson stars decreases until it reaches a minimum at the point denoted by A. The mass then increases again, crossing point B, which denotes the critical point of a first order phase transition \cite{Huang:2025dgc}. Since the scalarized solutions possess a higher binding energy beyond point B, they become the physically preferred solutions. The subfigure (b) exhibits the dimensionless mass versus the boson frequency of the GR (blue) and STT (red) solutions. 
In the upper right rectangle the STT solutions are the preferred solutions

\subsection{Thick accretion disks}

We now briefly recall thick accretion disks, which are composed of a perfect fluid, that surrounds a compact gravitating object \cite{Abramowicz:1978,Kozlowski:1978}. A stationary and axially symmetric metric is assumed for the compact object, which can be written in general form as
\begin{align}
ds^2=g_{tt}dt^2+2g_{t\varphi}dtd\varphi+g_{rr}dr^2+g_{\theta\theta}d\theta^2+g_{\varphi\varphi}d\varphi^2.
    \label{eq:generalmetric}
\end{align}
Thick tori arise by assuming also stationarity and axisymmetry for the fluid. The fluid particles four-velocity $u^\mu$ can then be represented as
\begin{equation}
u^\mu=u^t(\eta^\mu+\Omega \xi^\mu),
\end{equation}
where $\eta^\mu = (1,0,0,0)$, $\xi^\mu = (0,0,0,1)$ and $\Omega$ is the angular velocity, $\Omega \dot{=} \frac{u^\varphi}{u^t}$.
The fluid four-acceleration $a_\mu$ can then be written as a function of the normalized mass energy $u_t$, the angular velocity $\Omega$ and the specific angular momentum $\ell$ of the fluid particles,
\begin{equation}
a_\mu=\partial_\mu \ln|u_t|-\frac{\Omega}{1-\Omega \ell}\partial_\mu \ell.
\label{ac}
\end{equation}
The right hand side gives one part of the relativistic Euler equations for a fluid, which relate thermodynamic to kinematic quantities. The other side is given in terms of the rest-mass density $\rho$, the specific enthalpy $h$ and the pressure $p$,
\begin{equation}
-\frac{1}{\rho h}\partial_\mu p=\partial_\mu\ln|u_t|-\frac{\Omega}{1-\Omega \ell}\partial_\mu \ell.
\label{euler}
\end{equation}
Integrability of the expression above is given when $\Omega = \Omega(\ell)$, which is ensured under the assumption of a barotropic equation of state (see e.g.~\cite{Rezzolla:2013dea}). By integrating the equation we can define an effective potential $\mathcal{W}$, which represents the potential and centrifugal energy of the fluid particles,
\begin{equation}
    \mathcal{W}(r,\theta)-\mathcal{W}_{in}=\ln{|u_t|}-\ln{|(u_t)_{in}|}-\int_{\ell_{in}}^\ell\frac{\Omega d\ell'}{1-\Omega \ell'} = -\int_{p_{in}}^p \frac{dp'}{\rho h}, \
    \label{W2}
\end{equation}
where the subscript $in$ denotes the inner edge of the disk. $\mathcal{W}_{in}$ will be set to zero for all feature computations, since it marks the boundary between unbounded and bounded orbits. By approximating the specific angular momentum distribution roughly as constant $\ell(r, \theta) = \ell_0$, the corresponding integral vanishes in the equation above. In real astrophysical scenarios the accretion disks have a non constant distribution. However, by choosing a uniform distribution the integral simplifies, without losing the qualitative disk properties, since the possible disk configurations depend mainly on the spacetime geometry. The effective potential can then be fully expressed in terms of the metric components as
\begin{align}
    \mathcal{W}(r,\theta)=\ln\left(\frac{g_{t\varphi}^2-g_{tt}g_{\varphi\varphi}}{g_{\varphi\varphi}+2\ell_0g_{t\varphi}+\ell_0^2g_{tt}} \right)^{\frac{1}{2}}.
    \label{potentialW}
\end{align}
For a given metric, $\ell_0$ then acts as a free parameter which determines the shape of the potential within the realm of possibilities of the spacetime background. As such, the pressure and density are also functions of the spacetime metric (through equation (\ref{W2})). Since the equi-pressure and isodensity surfaces coincide with the equi-potential surface, we will focus our study on the latter, as they fully determine the morphological properties of the possible disk solutions.
Consequently, the extrema of $\mathcal{W}$ also correspond to extrema of the density $\rho$ and the pressure $p$. A minimum of $\mathcal{W}$ is called a disk center, since the density and pressure are at a maximum here. It is thus of major interest, as it determines the main physical properties such as the luminosity of the disk. A maximum of $\mathcal{W}$ corresponds to a minimum of $\rho$ and $p$ and is called a disk cusp. The cusp is of central importance for accretion processes, since at the cusp the motion of the disk matter becomes unstable and falls onto the central object. At the disk center and the disk cusp the fluid moves geodesically, and the orbits represent Keplerian orbits with the Keplerian specific angular momenta $\ell_K^\pm(r)$,
\begin{equation}
    \ell_K^\pm = -\frac{g_{t\varphi}+g_{\varphi\varphi}\Omega_K^\pm}{g_{tt}+g_{t\varphi}\Omega_K^\pm},
\end{equation}
where the signs specify prograde ($+$) and retrograde ($-$) motion, and $\Omega_K^\pm$ the Keplerian angular velocity, which is given by 
\begin{equation}
    \Omega_K^\pm = \frac{-\partial_r g_{t\varphi}\pm\sqrt{(\partial_r g_{t\varphi})^2-\partial_r g_{tt}\partial_r g_{\varphi\varphi}}}{\partial_r g_{\varphi\varphi}}.
\end{equation}
In case of a uniform specific angular momentum distribution of the disk, the intersections $\ell_K^\pm(r) = \ell_0$ reveal the extrema of $\mathcal{W}$, and by that also the disk center and disk cusp.

\section{Polish Doughnuts around self-interacting Boson Stars in GR and STT}

We consider here non self-gravitating perfect fluid disks around boson stars with a quartic self-interaction potential. Previously, thick disks were considered only for boson stars without self-interactions \cite{Meliani:2015zta,Teodoro:2020kok}. The family of GR boson stars depicted in Fig.~\ref{fig1} is only studied up to the maximum mass configuration, since the configurations beyond the maximum mass are prone to collapse. We will especially also put focus on the region of the phase transition between GR and STT boson stars, which is marked in Fig.~\ref{fig1} by the point B. However, since GR and STT boson stars are given by distinct spacetime geometries, we have to define a common coordinate for illustration, in order to enable interpretability. As such, we here choose the circumference radius $R_c$ for the radial coordinate, since it has an invariant physical interpretation across different spacetime geometries. It is defined by

\begin{align}
    R_c(r,\theta) = \sqrt{g_{\varphi \varphi}(r,\theta)},
\end{align}

and it corresponds to the radius a circle with a circumference of $2 \pi R_c$ would have in flat spacetime. Consequently, we extend this coordinate usage later on also to two-dimensional cross section plots.

\begin{figure}[H]
\centering
\begin{subfigure}{.325\textwidth}
  \centering
  \includegraphics[width=\linewidth,  height= 0.75\textwidth]{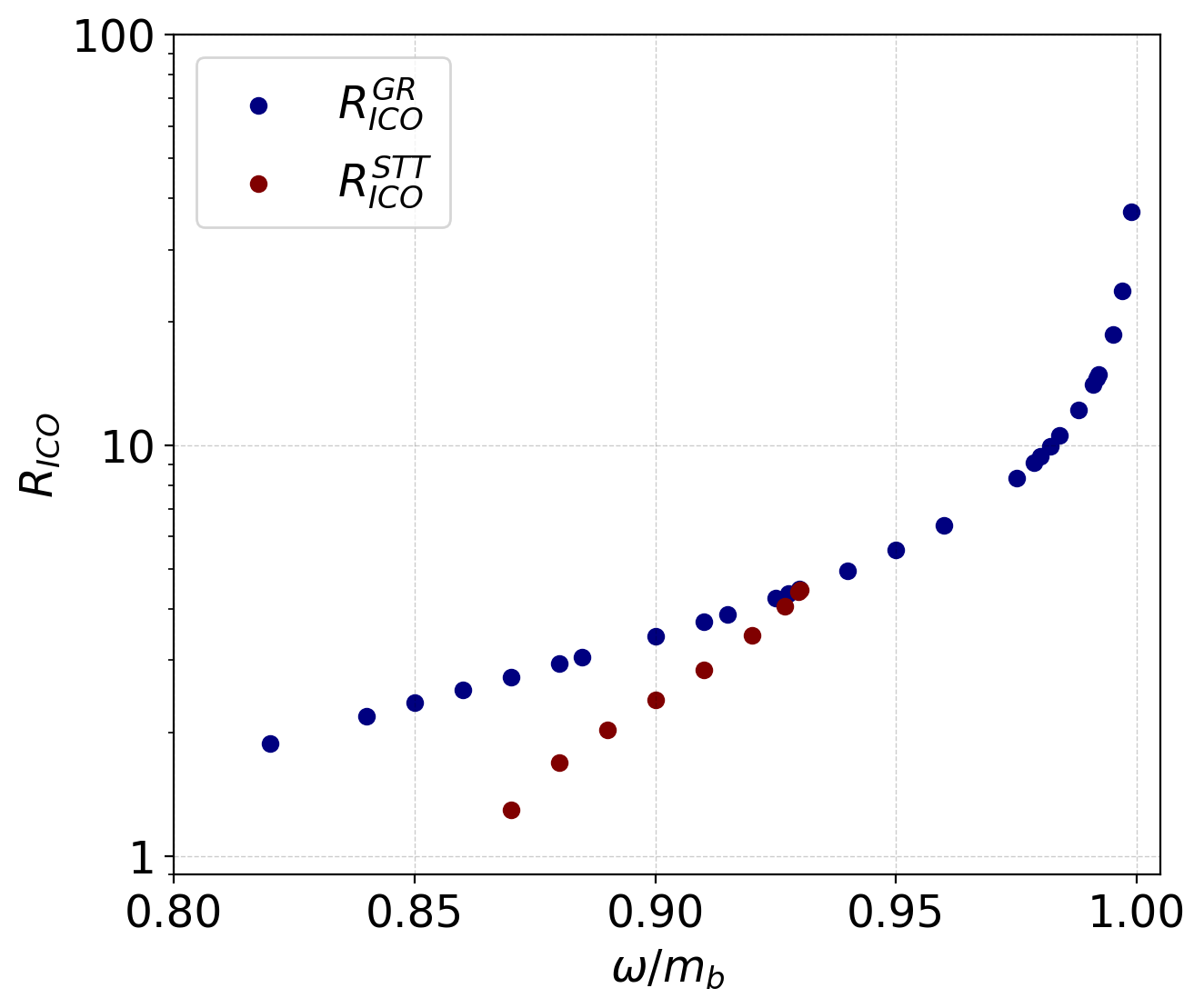}
  \caption{}
\end{subfigure}
\begin{subfigure}{.325\textwidth}
  \centering
  \includegraphics[width=\linewidth,  height= 0.75\textwidth]{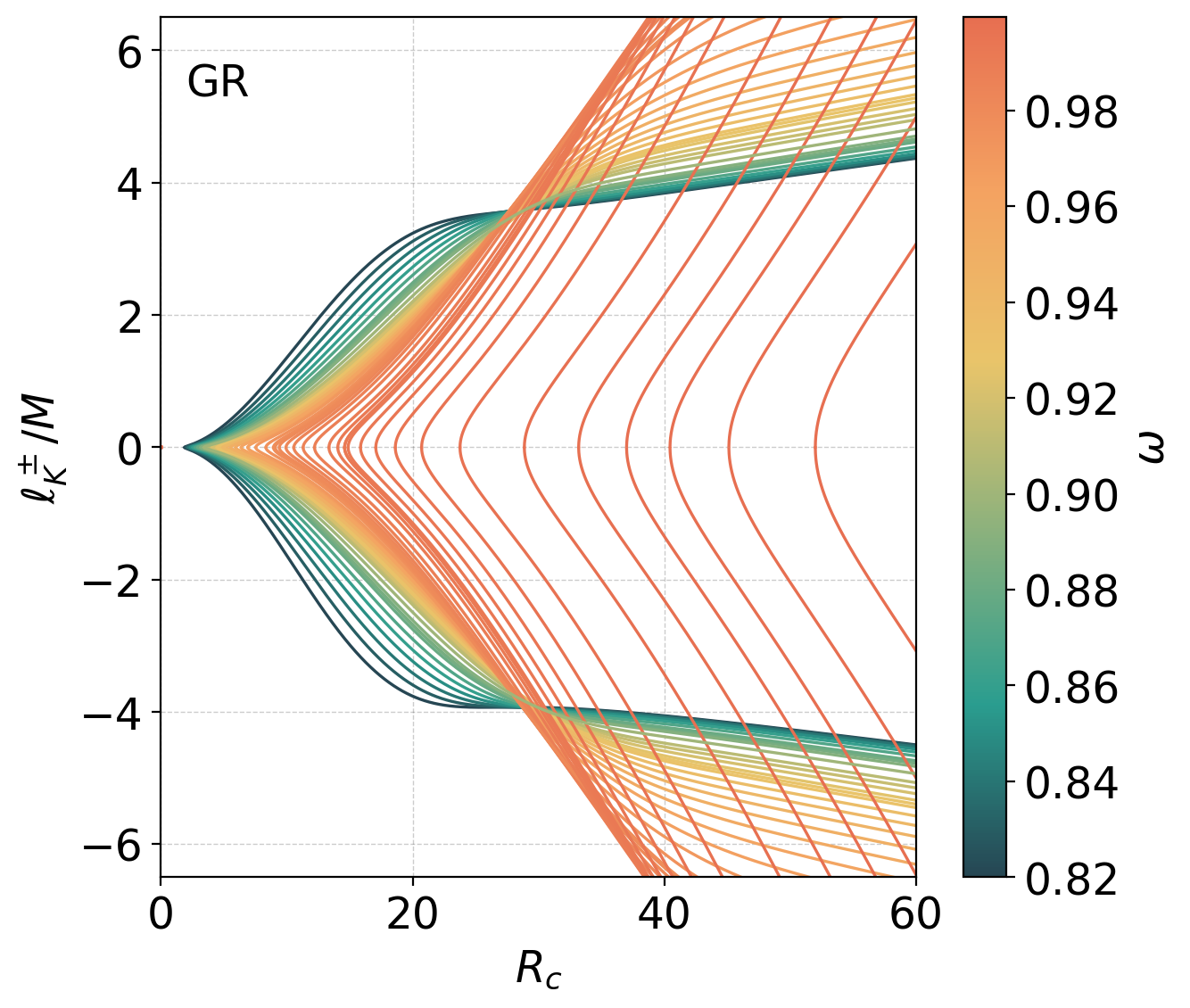}
  \caption{}
\end{subfigure}
\begin{subfigure}{.325\textwidth}
  \centering
  \includegraphics[width=\linewidth,  height= 0.75\textwidth]{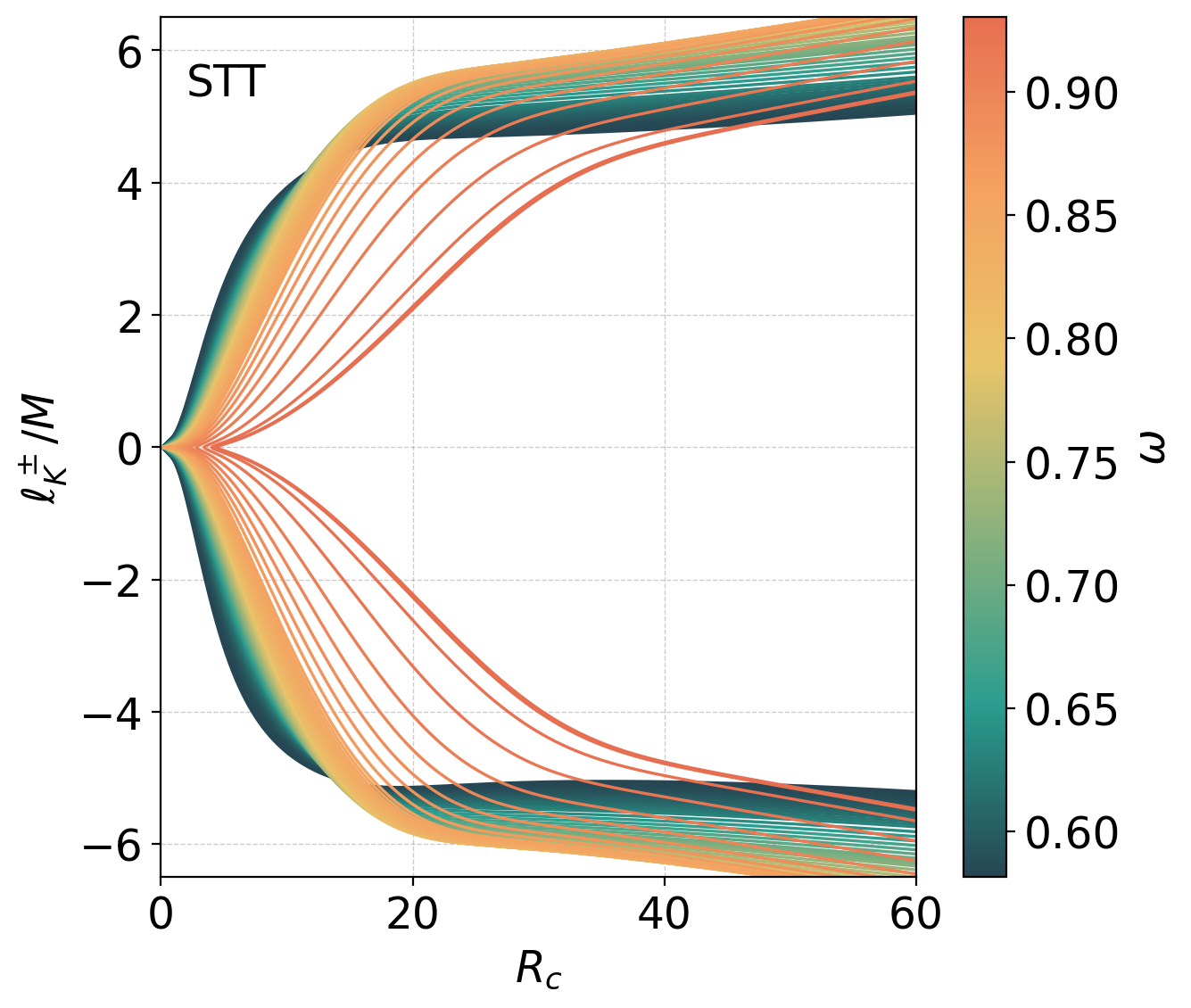}
  \caption{}
\end{subfigure}
\caption{(a) showcases the location of the ICO over the angular frequency in terms of the circumference radius for the different GR and STT boson stars. The point where both solutions merge corresponds to point C in Fig. \ref{fig1}. (b) and (c) showcase the scaled Keplerian specific angular momentum distribution in the equatorial plane for the GR and STT solutions. The different distributions are colored corresponding to their solutions $\omega$ according to the color scale next to the plots.}
\label{fig2}
\end{figure}

One property often encountered in boson star spacetimes is a minimal radius for geodesic circular orbits, a so-called \textit{innermost circular orbit} (ICO), whose radial value $r_{ICO}$ depends on the given boson star metric \cite{Grandclement:2014msa,Meliani:2015zta}.
The ICO marks the minimum radius up to which circular geodesics exist.
We found for self-interacting boson stars, that at this radius the angular momentum on the prograde and retrograde circular geodesics become degenerate with respect to a zero angular momentum observer (ZAMO). If an ICO is present, it marks by definition the innermost possible location of a disk center, as no circular geodesic motion is possible closer to the coordinate origin. In Fig.~\ref{fig2} we showcase the location of the ICO for the GR and STT boson stars, as well as the specific angular momentum distribution for the space of GR and STT solutions.

As a qualitative difference between GR and STT boson stars, we here identify the existence of an ICO across the solution space. For both branches the ICO is a monotonically increasing function of $\omega$, however, for the STT boson stars it only exists for the upper $\omega$ range of solutions, whereas for the GR boson stars an ICO is present for all solutions. Thus, for the lower part of the STT solution space, stable circular orbits can exist up to the origin. As such, circular orbits could sit at every location in the equatorial plane across the spacetime. This enables the formation of accretion disks, which lie close to the symmetry axis and are morphologically closer to ellipsoids than to tori (topologically they are nonetheless still a toroid, with the exception of the limiting case for vanishing disk momentum that we explore further down). For the upper $\omega$ part of STT solutions, the location of the ICOs approaches each other as $\omega$ increases, and it merges at point C (Fig.~\ref{fig1}). Beyond point C, the location of the ICO shifts strongly outwards at the upper $\omega$ end of GR solutions.

The Keplerian specific angular momenta are for the GR branch strictly increasing functions of the radial coordinate, which implies that all circular orbits are also of stable nature. In case of the STT solutions there is a range of solutions, $\omega \leq 0.602$, for which $\ell_K^-$ is non-monotonic, it has a local minimum and a local maximum. The locations of these extrema mark marginally stable retrograde orbits. The region of stability reaches inwards from the local minimum and outwards from the local maximum. Thus, in the region between those extrema, all retrograde circular geodesics are of unstable nature. This indicates the possibility of two-centered accretion disks structures, which represent a significant qualitative difference with regard to the GR solutions. In these two-centered solutions an outer pressure center could be linked through a cusp to an inner pressure center. The inner center is hereby located in the inner region of stability, the outer center is located in the outer region of stability, and the cusp is located in the region of instability between them. The flow at the cusp is geodesic but unstable, which naturally implies accretion processes, where matter is transferred from the outer regions inwards through the cusp towards the inner regions.

A further notable property of both branches lies in the continuity of the Keplerian specific angular momentum distribution, which enables particles on geodesics that can have a specific angular momentum of zero and arbitrarily close values to zero. In case of the GR solutions, the circular geodesics with $\ell_K  = 0$ correspond always to the ICO, which is also the case for the STT solutions that have an ICO. For the STT solutions which do not have an ICO, circular orbits approach $\ell_K  = 0$ as they get closer to the origin, with $\ell_K^\pm(r=0)=0$. Thus, Polish Doughnuts which have a center close to the origin have therefore also a rather small specific angular momentum. As a consequence, they exhibit a more spherical morphology, since the influence of the centrifugal forces decreases, and the biggest fraction of the pressure balance is contributed by a hydrostatic equilibrium of gravitational attraction and thermodynamic pressure. For some of the GR solutions such spherically appearing disks are even possible for configurations where the disk center lies far from the origin, since in the upper $\omega$ range of solutions the ICO is located far away from the origin.

\begin{figure}[H]
\centering
\begin{subfigure}{.327\textwidth}
  \centering
  \includegraphics[width=\linewidth,  height= 0.75\textwidth]{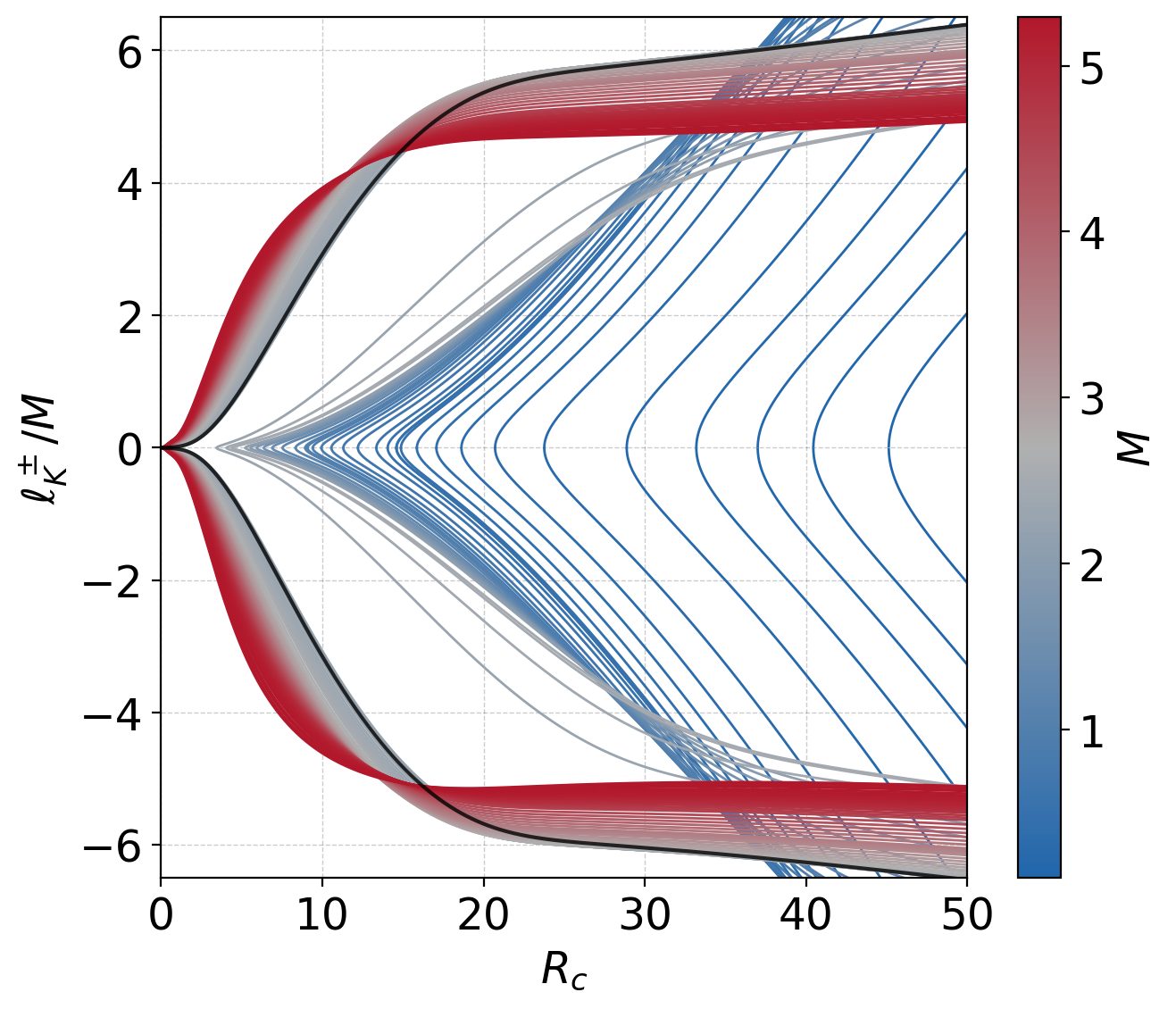}
  \caption{}
\end{subfigure}
\begin{subfigure}{.327\textwidth}
  \centering
  \includegraphics[width=\linewidth,  height= 0.75\textwidth]{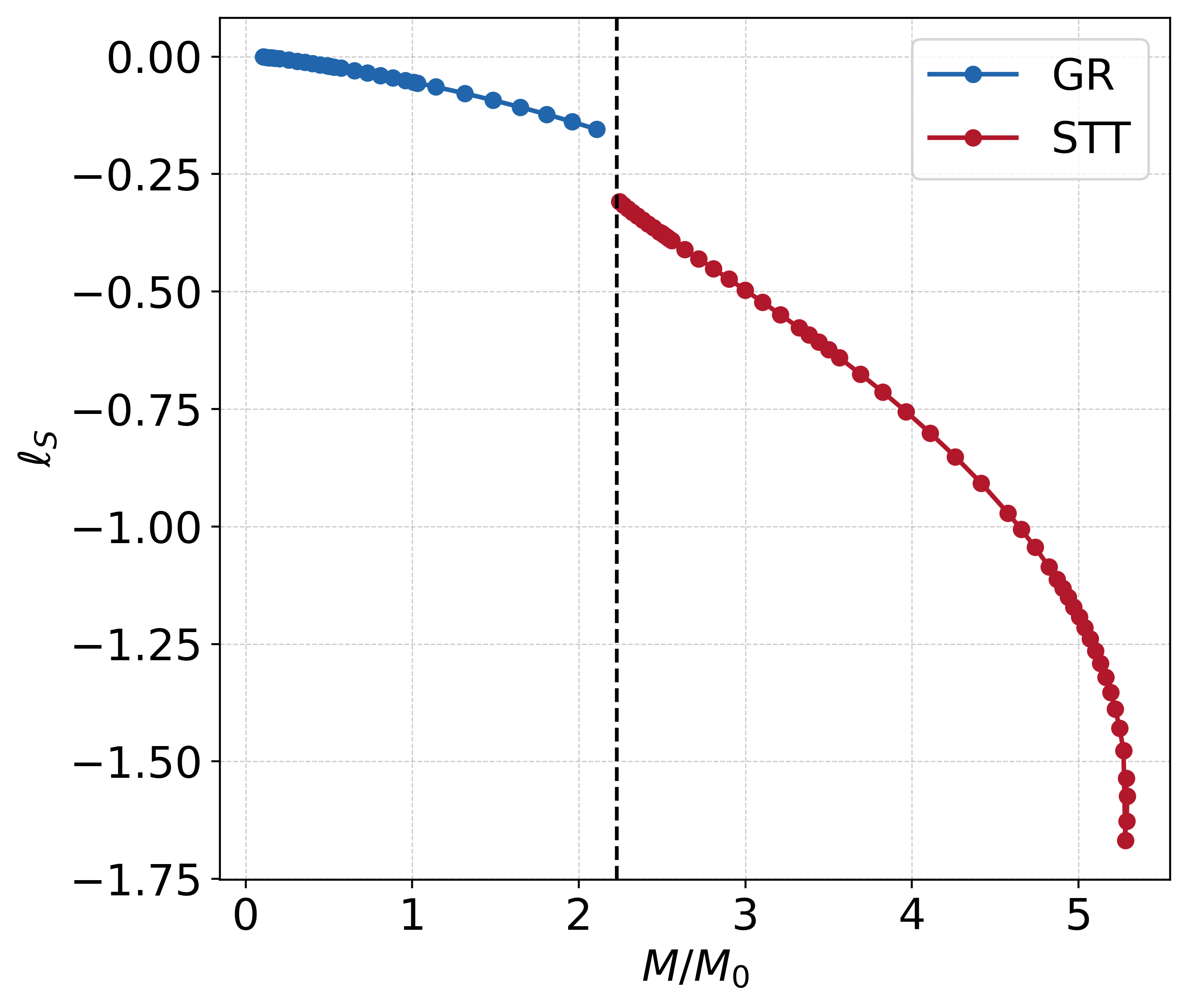}
  \caption{}
\end{subfigure}
\begin{subfigure}{.327\textwidth}
  \centering
  \includegraphics[width=\linewidth,  height= 0.75\textwidth]{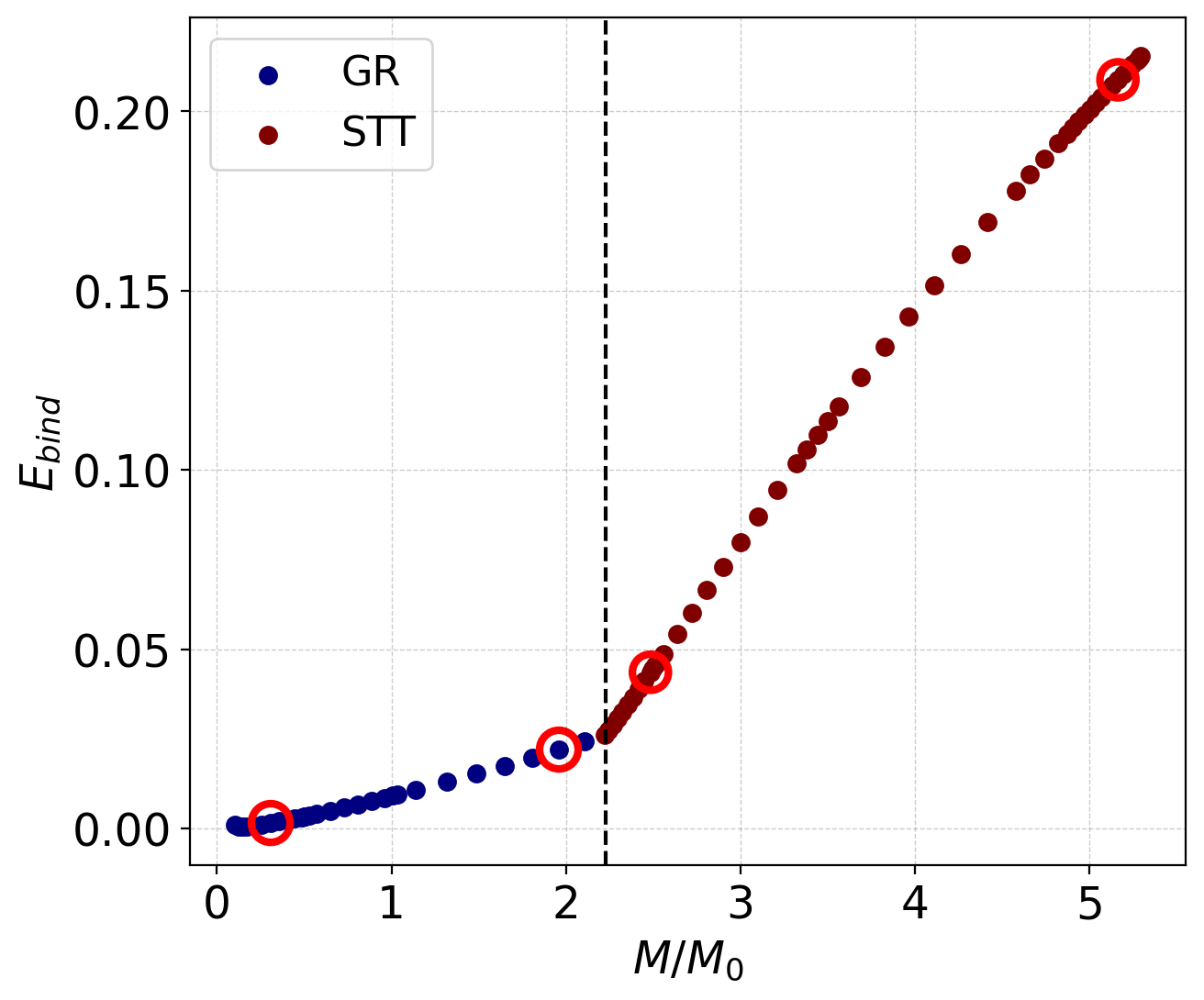}
  \caption{}
\end{subfigure}
\caption{(a) $\ell_K^\pm$ distribution for the energetically preferred solutions. The black curve marks the transition from GR to STT solutions. (b) lower limit $\ell_S$ of the disk specific angular momentum up to which static orbits and thus static surfaces exist within disk solutions, plotted over the mass. The range is then given by $\ell_0 \in (\ell_S,0)$. The dashed vertical line marks the transition from GR to STT. (c) binding energy over the mass for the energetically preferred solutions. The red circles mark the solutions which were selected for a representative analysis of disk solutions (GR: $M \in \{0.308, \ 1.960\}$, STT: $M \in \{2.484, \ 5.163
\}$).}
\label{fig3}
\end{figure}

Furthermore, since all the $\ell_K$ curves approach zero, also static orbits are present across the whole solution spectrum. Static orbits are orbits where a test particle has zero angular velocity in the rest frame of the spacetime \cite{Collodel:2017end}. Thus, the only contribution to its angular momentum for a ZAMO, originates then from the frame dragging of the spacetime. These orbits are defined in the equatorial plane by
\begin{align}
    \ell_0 = \ell_r(r) \coloneqq - \frac{g_{t \varphi}}{g_{tt}} \Bigg|_{\theta=\frac{\pi}{2}},
\end{align}
where $\ell_0$ is the angular momentum of the disk particles (which is for Polish Doughnuts constant across the disk). In the presence of static orbits \textit{static surfaces} arise for retrograde tori, which represent the generalization of static orbits beyond the equatorial plane. Static surfaces are toroidal surfaces inside the disk fluid, where the fluid particles move on static orbits \cite{Teodoro:2020kok}. This enables tori, where the fluid particles change their rotational direction at the static surface with respect to the rotating rest frame. The lower limit for the specific angular momentum for which such static orbits and surfaces can exist for GR and STT boson stars is depicted in Fig. \ref{fig3} (b). The range of existence is here an increasing function of the mass. The phase transition from GR to STT boson stars marks a discontinuous shift, increasing the range for STT boson stars.

In the further analysis we focus only on the energetically preferred boson star solutions from the GR and STT branches (illustrated in Fig.~\ref{fig1}), since they would be the ones relevant for any astrophysical scenario. The figures \ref{fig3} (a) and (c) depict the spectrum of these solutions with regard to $\ell_K^\pm$ and the binding energy of the boson stars.

\begin{figure}[H]
    \centering
    \mbox{
        \includegraphics[width=\linewidth]{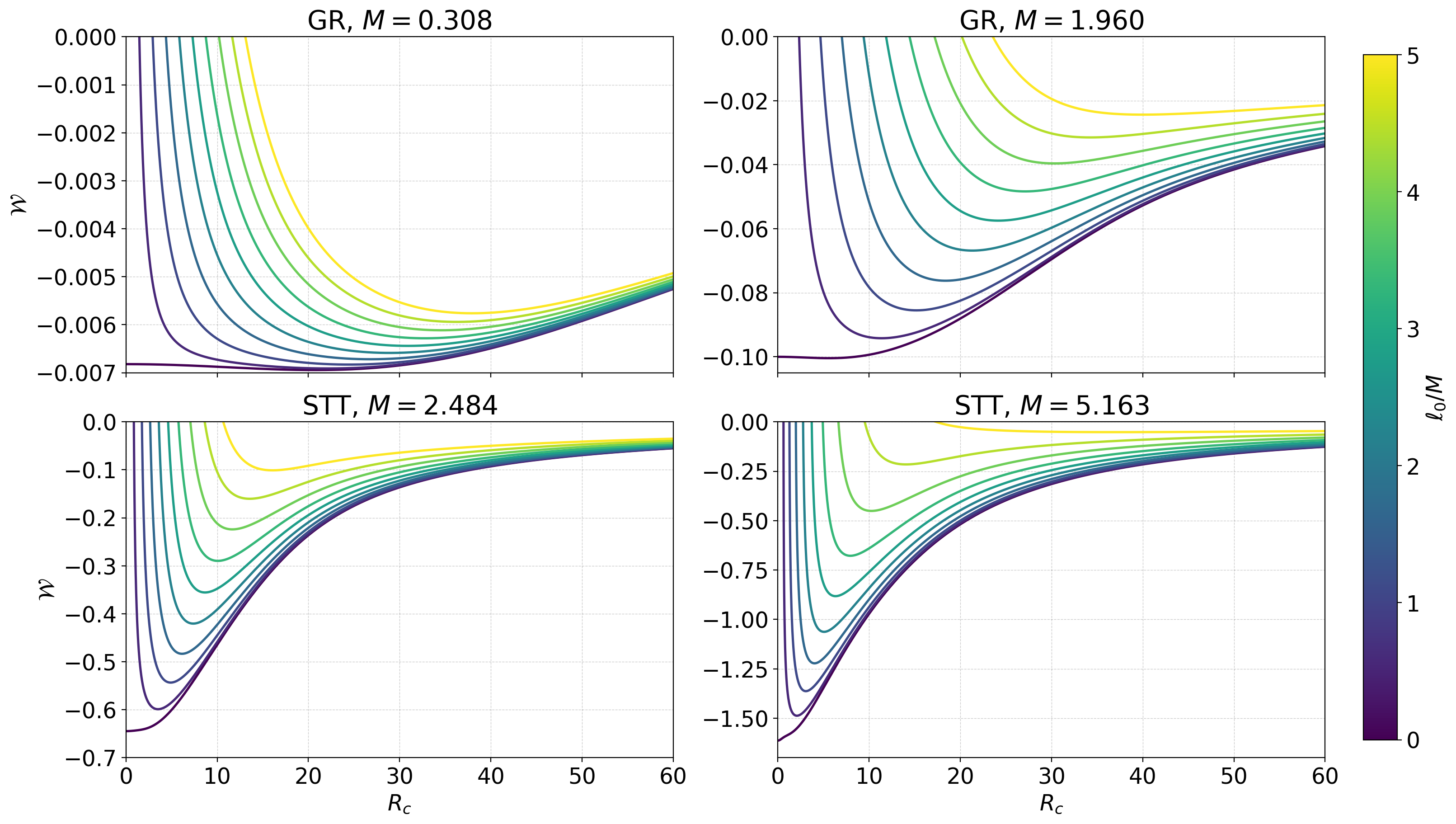}}
    \caption{Equatorial effective potential for different prograde disk solutions for the picked GR and STT boson stars. The color scale represents solutions for different values of the specific angular momentum of the disks. The lowest curve in each subplot represents the special solution with $\ell_0 = 0$.}
    \label{fig4}
\end{figure}

For a representative analysis we picked here four different solutions, two from the GR branch and two from the STT branch. Each of the solutions represents either the upper/lower end of the solution space or the region close to the transition boundary. Since all the boson star solutions within each branch lie on a smooth continuum, the properties of solutions in between the picked representatives vary accordingly in a monotonic manner as far as we observed. The GR solutions with smaller mass allow only thick tori, where the disk center is located far away from the origin, since the ICO is an inversely increasing function of $M$. The choice of the specific angular momentum has here only small impact on the tori within this solution range (depicted in Fig.~\ref{fig4} upper left). The effective potential depth of the thick tori is naturally an increasing function of the mass for fixed $\ell_0$, where the most compact disks exist for the most massive solutions, which sit on the STT branch. The most compact disks of the high mass STT branch have a ~220 times deeper potential depth than the disks of the small mass GR branch (Fig.~\ref{fig4} upper left and lower right). Moreover, the differences between GR and STT boson stars are even significant close to the transition point, despite the comparable masses. This can be explained by the gap in the specific angular momentum distribution between the last GR and the first STT solution (Fig.~\ref{fig3} (a)). Accretion tori around the STT boson stars have deeper potential minima for the same scaled specific angular momenta, which are also located  closer to the origin, and in general, they are more affected by the chosen value of the specific angular momentum.

\begin{figure}[H]
\centering
\begin{subfigure}{.4\textwidth}
  \centering
  \includegraphics[width=\linewidth,  height= 0.75\textwidth]{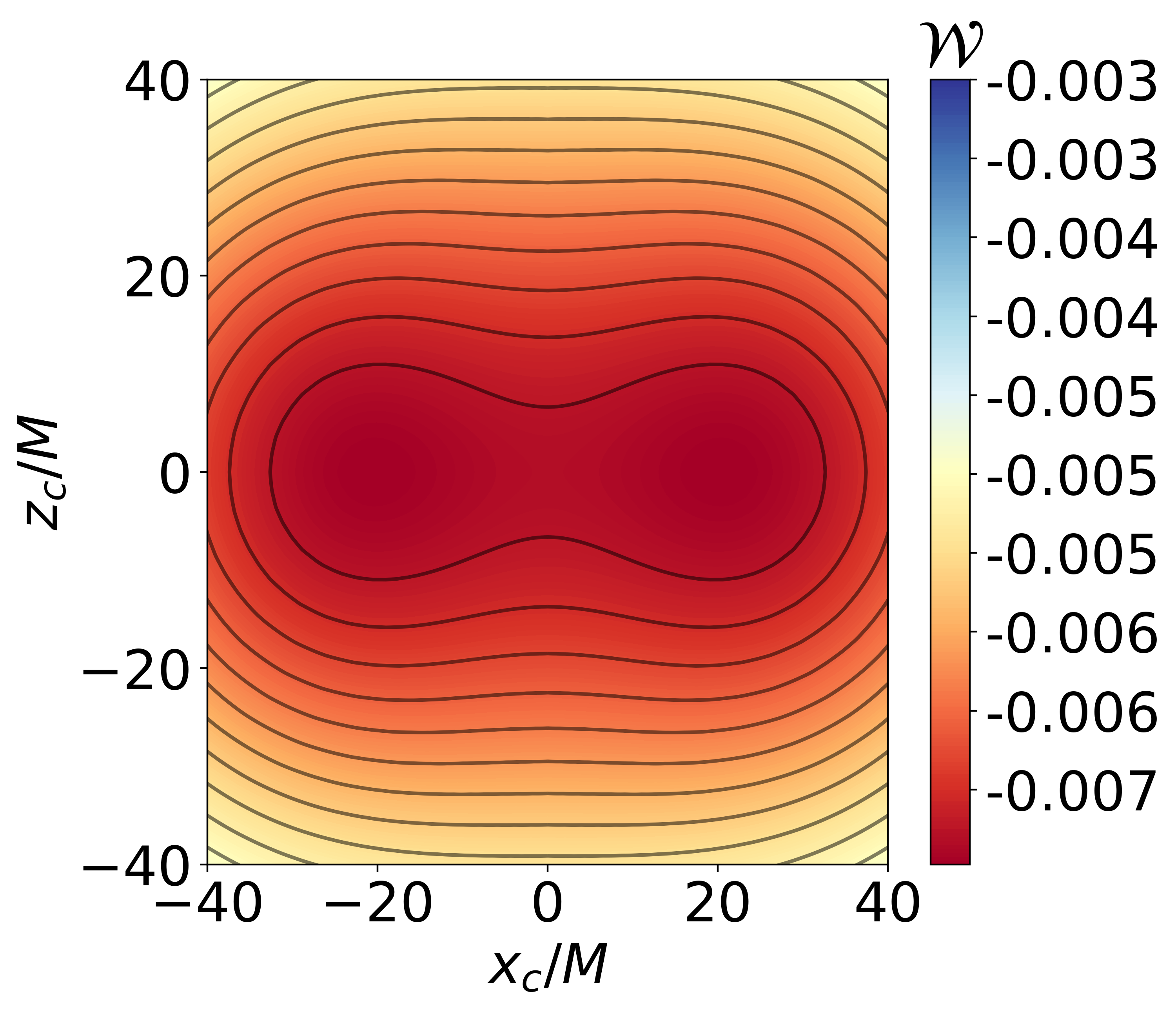}
  \caption{}
\end{subfigure}
\begin{subfigure}{.4\textwidth}
  \centering
  \includegraphics[width=\linewidth,  height= 0.75\textwidth]{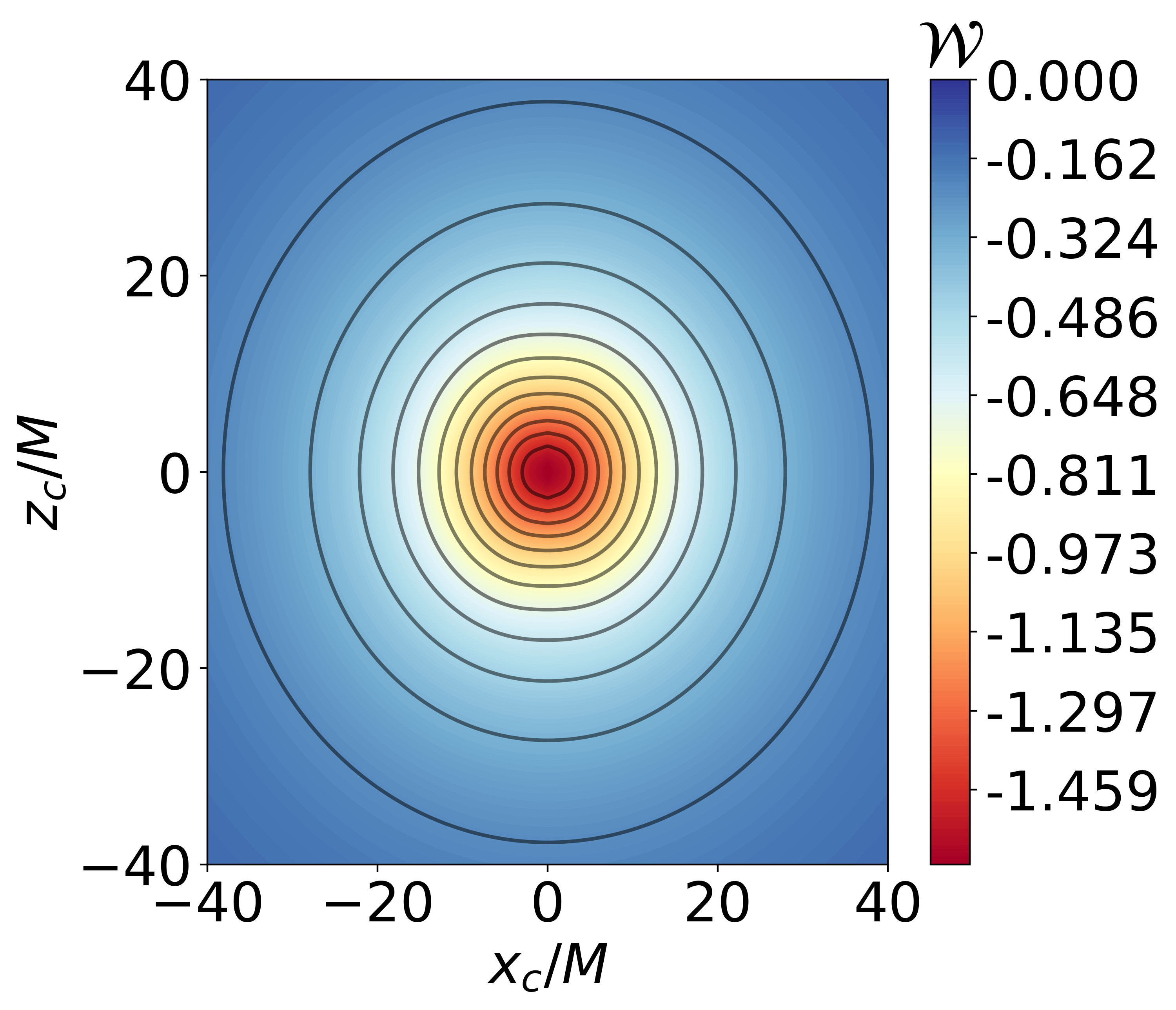}
  \caption{}
\end{subfigure}
\caption{(a) cross section plot of the disk effective potential with $\ell_0 = 0$ for the GR boson star with the $M = 0.308$. The $x$ and $z$ axes are given by pseudo-cylindrical coordinates using the circumference radius $R_c$ as the radial coordinate. (b) the same for the STT boson star with $M = 5.163$.}
\label{fig5}
\end{figure}

The lowest curve in each plot in Fig.~\ref{fig4} represents the torus solution with vanishing angular momentum. In these solutions the disk center is located either at the ICO, $r_{center} = r_{ICO}$ (Fig.~\ref{fig4} upper row) or at the coordinate origin if no ICO is present, $r_{center} = 0$ (Fig.~\ref{fig4} lower row). In these zero angular momentum solutions the tori extend all the way to the origin, where either a cusp or a center is located, depending on the presence or absence of an ICO. The topology of the disk changes from toroidal to ellipsoidal. This follows from the absence of centrifugal forces, which leaves only the gravitational pull and the thermodynamic pressure as the acting forces, leading to a connected structure in hydrostatic equilibrium. As depicted in Fig.~\ref{fig5}, the equi-potential surfaces around the center region are dumbbell-shaped for less massive (GR) boson star solutions, with a rather uniform potential gradient distribution across the disk. For more massive (STT) boson stars, the disk exhibits a higher degree of spherical symmetry with steep potential gradients. Most of the disk mass here is concentrated in a narrow region around the origin, creating a highly compact sphere. The zero angular momentum tori can be viewed as the asymptotic limit of more physically realistic small angular momentum tori and can thus act as an approximation of their general properties.

\begin{figure}[H]
\centering
\begin{subfigure}{.4\textwidth}
  \centering
  \includegraphics[width=\linewidth,  height= 0.75\textwidth]{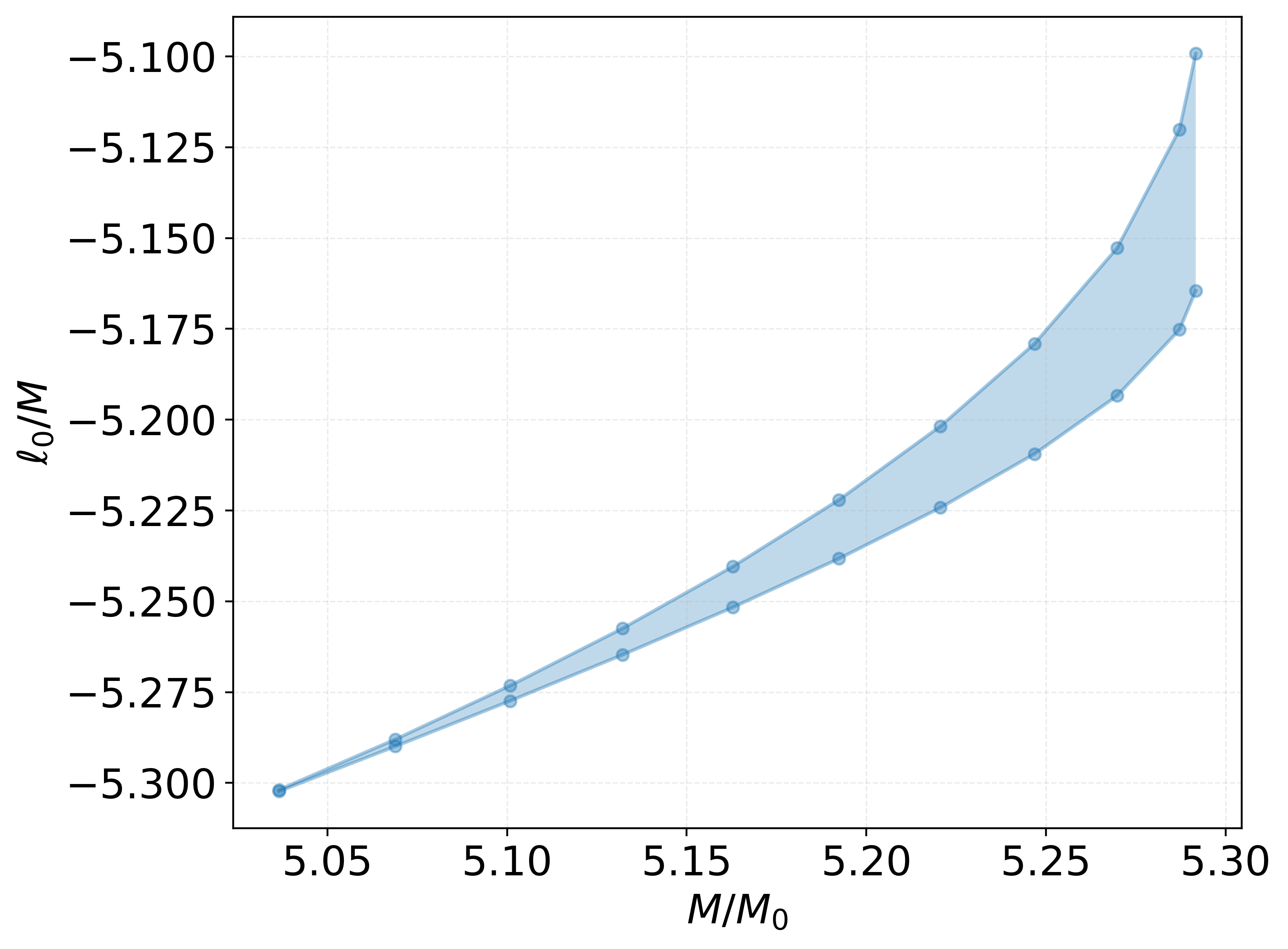}
  \caption{}
\end{subfigure}
\begin{subfigure}{.4\textwidth}
  \centering
  \includegraphics[width=\linewidth,  height= 0.75\textwidth]{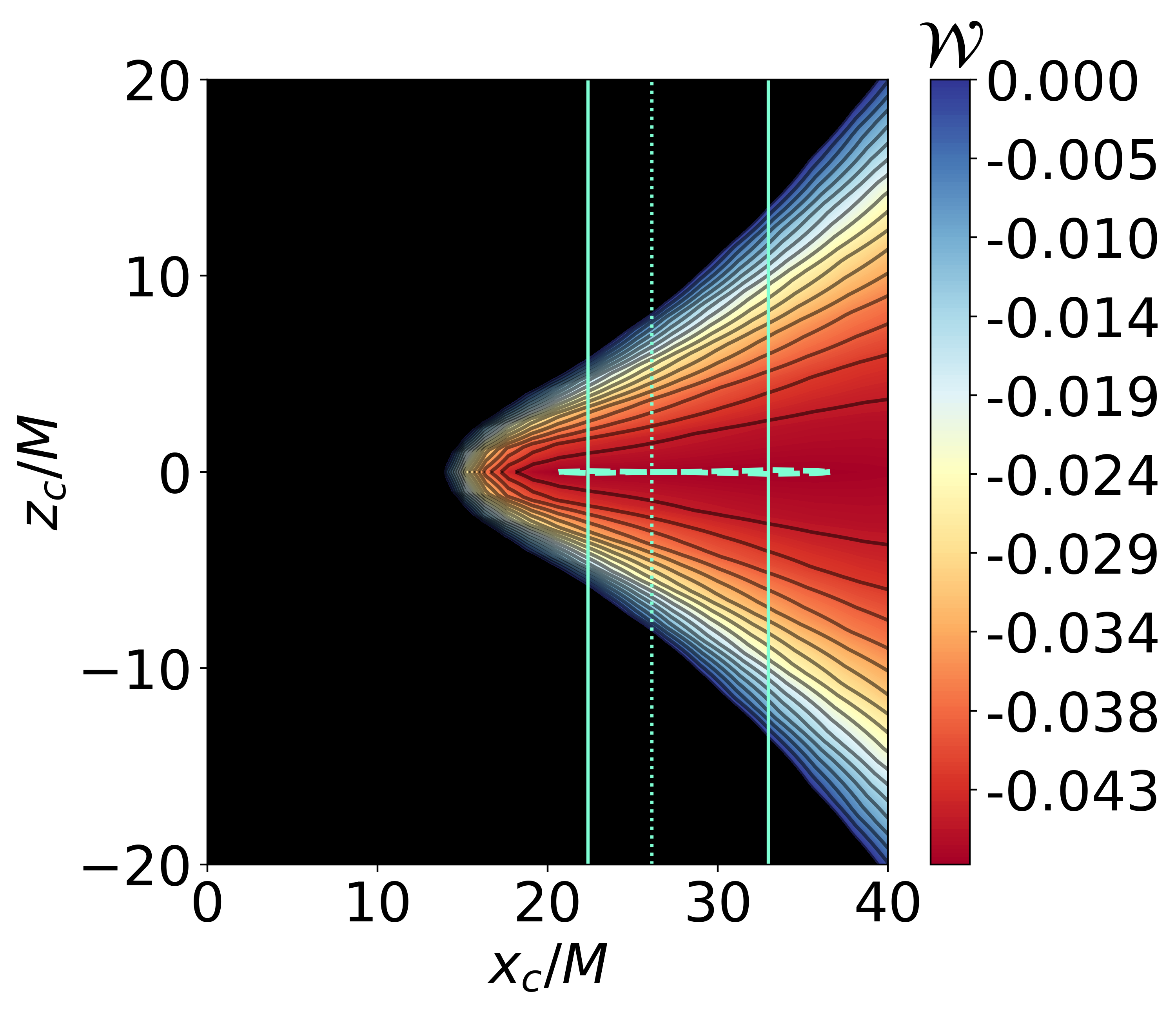}
  \caption{}
\end{subfigure}
\caption{(a) range of $\ell_0$ for which two-centered disks exist for the most massive STT boson stars, illustrated by the shaded area. (b) exemplary cross section plot for a two-centered disk around the STT boson star with $M = 5.163$ with $\ell_0 = -5.252M$. The solid vertical lines mark the inner and outer center, the dotted vertical line marks the location of the cusp. The colored horizontal curve marks the equi-potential surface of the cusp, which has only a small vertical extend, since the difference between the potential values of both centers and the cusp is small.}
\label{fig6}
\end{figure}

The presented solutions in Fig.~\ref{fig4} showcase only prograde tori; in the case of retrograde tori, we found that the behavior is mostly similar, and the difference in the general disk properties is not qualitatively significant. However, there is one important exception, which is present for the most massive solutions of the STT branch ($M > 5.036$). For these boson stars, there exists a narrow range for $\ell_0$, for which two-centered disk solutions exist. In these two-centered solutions an outer center is connected via a cusp to an inner center (depicted in Fig.~\ref{fig6} (b)). Albeit only representing a narrow range of the solution space, the existence of such two-centered disks presents one major qualitative difference between GR and STT boson stars. Such two-centered solutions were in the past also found for boson stars in GR without a self-interaction potential \cite{Meliani:2015zta,Teodoro:2020kok}. It should be noted, that the parameter range for two-centered solutions may increase in more generalized thick disks models, such as magnetized disks or configurations with a non-constant angular momentum distribution.

\section{Conclusions}
We analyzed Polish Doughnuts around rotating boson stars with a self-interacting potential in general relativity and their spontaneously scalarized counterparts in scalar tensor theory. Here, we focused only on the branches which represent energetically preferred solutions, since they should play the most relevant role for astrophysical scenarios. For a qualitative disk analysis, we studied the Keplerian specific angular momentum of the solutions and found that all GR boson stars possess an innermost circular orbit marked by a vanishing angular momentum. In contrast, for the more massive scalar-tensor theory solutions, no innermost circular orbit exists, and orbits are present all the way up to the origin. Almost all Keplerian specific angular momentum distributions are strictly monotonically increasing except for the most massive scalar-tensor theory boson stars, where retrograde orbits possess a small region of instability between an inner and outer stable region. This enables two-centered disk solutions, which were already known for boson stars with no self-interaction in general relativity, where the inner and outer disk regions are connected via a cusp. Furthermore, there also exists a small range for disk solutions across both boson star branches, where static orbits within the disk exist. This range increases with increasing mass of the solutions. Comparing the general relativity and scalar-tensor theory disks with each other reveals a strong influence of the scaled respective mass. The general relativity disk solutions tend to have a more uniform density distribution and are less energetically bound, whereas the more massive STT disk solutions represent highly compact objects, with the largest amount of mass closely distributed around the origin. This also applies to the special solutions with vanishing angular momentum, where the disk is dumbbell shaped for the less massive boson stars, whereas for the more massive ones, the disk possesses a high degree of spherical symmetry around the origin. In general, the less massive boson star solutions within the GR branch tend to be less sensitive to different values of the disk angular momentum compared to their STT counterparts.

We conclude that the additional terms introduced by the scalar-tensor theories have indeed a qualitative effect across the solution space of boson stars when it comes to Polish Doughnuts. These differences are mainly attributed to orbital dynamics and properties of the ICO, as well as the more massive nature of the energetically preferred scalar-tensor boson star solutions.

\section*{Acknowledgement}
P.N. gratefully acknowledges support from the Bulgarian National Science Fund (NSF) under Grant KP-06-DV/8 within the funding programme “VIHREN–2024”.

\end{document}